\documentclass[aps,showpacs,longbibliography,notitlepage,superscriptaddress,twocolumn]{revtex4-1}
\pdfoutput=1
\usepackage[utf8]{inputenc}
\usepackage[english]{babel}
\usepackage[T1]{fontenc}
\usepackage{amsmath}
\usepackage{xcolor}
\colorlet{myPurple}{blue!40!red}
\colorlet{myPurplee}{blue!10!red}
\colorlet{myCyan}{cyan!60!gray}
\colorlet{myRed}{red!66!black}
\usepackage{tikz}
\usepackage{pgfplots}
\pgfplotsset{compat=1.14}
\usepackage[colorlinks=true,citecolor=myRed,urlcolor=myRed,linkcolor=myRed]{hyperref}
\usepackage[normalem]{ulem}
\usepackage{exscale}
\usepackage{bbm}
\usepackage{graphicx}
\usepackage{amsmath}
\usepackage{latexsym}
\usepackage{amsfonts}
\usepackage{amssymb}
\usepackage{times}
\usepackage[T1]{fontenc}
\usepackage{amsthm}
\usepackage{enumerate}
\usepackage{bbold}
\usepackage{color}
\usepackage{nicefrac}
\newcommand{\sket}[1]{{\ensuremath{\lvert#1\rangle}}}
\newcommand{\lket}[1]{{\ensuremath{\left\lvert#1\right\rangle}}}
\newcommand{\ket}[1]{\if@display\lket{#1}\else\sket{#1}\fi}
\newcommand{\tp}{\otimes}
\newcommand{\sbra}[1]{{\ensuremath{\langle#1\rvert}}}
\newcommand{\lbra}[1]{{\ensuremath{\left\langle#1\right\rvert}}}
\newcommand{\bra}[1]{\if@display\lbra{#1}\else\sbra{#1}\fi}

\newcommand{\sbraket}[2]{{\ensuremath{\langle#1\rvert#2\rangle}}}
\newcommand{\lbraket}[2]{{\ensuremath{\left\langle#1\!\left\rvert\vphantom{#1}#2\right.\!\right\rangle}}}
\newcommand{\braket}[2]{\if@display\lbraket{#1}{#2}\else\sbraket{#1}{#2}\fi}

\newcommand{\sketbra}[2]{{\ensuremath{\lvert #1\rangle\!\langle #2\rvert}}}
\newcommand{\lketbra}[2]{{\ensuremath{\left\lvert #1\right\rangle\!\!\left\langle #2\right\rvert}}}
\newcommand{\ketbra}[2]{\if@display\lketbra{#1}{#2}\else\sketbra{#1}{#2}\fi}
\usepackage{tcolorbox}
\usepackage{mathtools}

\newcommand{\al}{\alpha}
\newcommand{\gm}{\gamma}

\newcommand{\sx}{\sigma_{\mathsf{x}}}
\newcommand{\sz}{\sigma_{\mathsf{z}}}

\newcommand{\tr}{\textrm{Tr}}

\newcommand{\idd}{\mathds{1}}

\newcommand{\rA}{\text{A}}
\newcommand{\rB}{\text{B}}
\newcommand{\rC}{\text{C}}

\usepackage{tikz}
\usepackage{lipsum}
\theoremstyle{plain}
\newtheorem{thm}{Theorem}

\usepackage{graphicx}
\usepackage{bm}
\usepackage{dsfont}
\usepackage{tikz}
\usepackage[T1]{fontenc}
\usepackage{amsthm}
\usepackage{array}
\usepackage{amssymb}
\usepackage{amsfonts}
\usepackage{cancel}
\usepackage[toc,page]{appendix}
\usepackage{multirow}
\usepackage{color}
\usepackage{calrsfs}
\usetikzlibrary{backgrounds,decorations.pathreplacing,calc}

\usepackage{tkz-euclide}

\newcommand{\ivan}[1]{{\color{purple} #1}}
\newcommand{\jd}[1]{{\color{cyan} #1}}

\DeclareMathAlphabet{\mathcal}{OMS}{cmsy}{m}{n}

\begin{document}

\author{Ivan \v{S}upi\'{c}}
\email{ivan.supic@lip6.fr}
\affiliation{Département de Physique Appliquée, Université de Genève, 1211 Genève, Switzerland}
\affiliation{CNRS, LIP6, Sorbonne Universit\'{e}, 4 place Jussieu, 75005 Paris, France}
\author{Jean-Daniel Bancal}
\affiliation{Département de Physique Appliquée, Université de Genève, 1211 Genève, Switzerland}
\affiliation{Universit\'{e} Paris-Saclay, CEA, CNRS, Institut de physique th\'{e}orique, 91191, Gif-sur-Yvette, France
}
\author{Yu Cai}
\affiliation{Département de Physique Appliquée, Université de Genève, 1211 Genève, Switzerland}
\author{Nicolas Brunner}
\affiliation{Département de Physique Appliquée, Université de Genève, 1211 Genève, Switzerland}

\title{Genuine network quantum nonlocality and self-testing}

\date{\today}

\begin{abstract}
    The network structure offers in principle the possibility for novel forms of quantum nonlocal correlations, that are proper to networks and cannot be traced back to standard quantum Bell nonlocality. Here we define a notion of genuine network quantum nonlocality. Our approach is operational and views standard quantum nonlocality as a resource for producing correlations in networks. We show several examples of correlations that are genuine network nonlocal, considering the so-called bilocality network of entanglement swapping. In particular, we present an example of quantum self-testing which relies on the network structure; the considered correlations are non-bilocal, but are local according to the usual definition of Bell locality.
\end{abstract}

\maketitle

\section{Introduction}

Discovered by Bell in the 1960s, the phenomenon of quantum nonlocality has far reaching consequences \cite{Bell,review}. Arguably among the most counterintuitive aspects of quantum theory, it also represents a powerful resource for information processing, notably in the context of device-independent protocols~\cite{Acin2007,Pironio2010,Colbeck}. 

Quantum nonlocality manifests as strong correlations observed by distant parties performing local measurements on a shared quantum system. This phenomenon relies on the shared system to be in an entangled state, and the local measurements to be incompatible (for each party). In this sense nonlocality is based on the judicious combination of several inherently non-classical properties of quantum mechanics, namely entanglement of quantum states and incompatibility of quantum measurements. 

Recently, growing interest has been devoted to the exploration of quantum nonlocality in networks; see e.g. \cite{Armin} for a recent review. A general framework has been developed for this problem \cite{Branciard_2010,Fritz_2012,branciard2012bilocal}. Consider a network with several parties, and $N$ sources distributing physical (classical or quantum) systems to various subsets of parties. The main idea consists in assuming that all the sources of the network are independent from each other. This allows one to define the notion of $N$-locality, which can be seen as a natural extension of the concept of Bell locality to networks with independent sources. Correlations that do not admit an $N$-local model are termed ``network nonlocal''. Importantly, whether a given distribution is network nonlocal or not depends on the topology of the considered network. In particular, for certain networks there are correlations that are provably network nonlocal, while the same correlation would be local when considering the usual Bell scenario, see e.g. \cite{branciard2012bilocal}.

Quantum theory allows for correlations in networks that are not $N$-local. The first examples of ``quantum network nonlocality'' have been derived for the scenario of entanglement swapping, now also referred to as ``bilocality'' \cite{Branciard_2010}. In turn, a more astonishing effect was discovered, namely that quantum network nonlocality can be demonstrated without the need for measurement inputs, i.e. each party performs a single fixed measurement \cite{Fritz_2012,branciard2012bilocal}. This would of course be impossible in the standard Bell scenario, and hence appears at first sight as a novel form of quantum nonlocality proper to networks.

A first example of such ``quantum nonlocality without inputs'' was proposed by Fritz~\cite{Fritz_2012}, for the so-called triangle network. The latter features three parties, each pair of them connected by a different source. After inspection, it turns out, however, that Fritz's example can be viewed as a clever mapping of the standard Bell CHSH scenario into the triangle network configuration \cite{Fraser,Supic}. 

A second example was more recently proposed by Renou and co-authors~\cite{renou_2019}, also for the triangle network. The authors argued, however, that their example is fundamentally different from that of Fritz, and could represent a form of quantum nonlocality genuine to networks, i.e. that cannot be mapped back to any standard Bell scenario. This conjecture relies on the observation that their quantum distribution was based on all three sources producing an entangled state (contrary to Fritz's example, where only one source needed to produce entanglement), and all parties performing an entangled measurement on their incoming subsystems (again, Fritz's example uses only separable measurements). 

At this point, the main question is thus whether there exists indeed a form of quantum nonlocality genuine to networks, and how to formalize this concept. This is precisely the goal of the present paper. We first propose a formal definition of genuine network nonlocality. Our definition is operational, and considers standard Bell nonlocality as a resource. Correlations that can provably not be obtained by these resources are therefore termed genuine network nonlocal. In turn, we highlight several examples of this phenomenon. In particular, we present a self-testing result for the bilocality network, where the observation of specific quantum correlation certifies (up to local isometries) the form of local measurements (including a joint entangled measurement), and the underlying entangled states. This example shows that self-testing can be enforced by the network structure, in the sense that the quantum distribution is non-bilocal but can provably not violate any standard Bell inequality. 
We conclude with a perspective on future possible developments of these ideas.


\section{Setting}

\begin{figure}[t]
    \centering
    \includegraphics[width = 0.9\columnwidth]{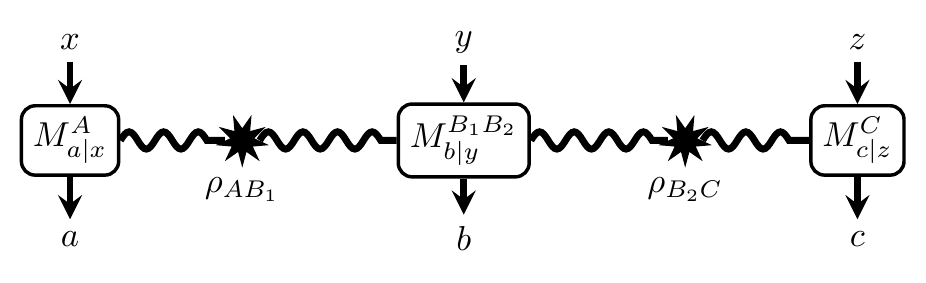}
    \caption{Bilocality (or entanglement swapping) network. 
    }
    \label{fig:bilocality}
\end{figure}



We focus our discussion on a simple tripartite quantum network, but our analysis can be straightforwardly extended to more general networks. The network we consider consists of three parties, Alice, Bob and Charlie, and two independent sources (see Fig.~\ref{fig:bilocality}). This network corresponds to the well-known entanglement swapping experiment \cite{Zukowski}. Alice and Bob share a bipartite quantum state $\rho_{AB_1}$, distributed by a common source $S_{AB}$. Similarly, Bob and Charlie share a state $\rho_{B_2C}$, distributed by a second source $S_{BC}$. 

Each party receives an input (i.e. a choice of measurement setting) and outputs the corresponding measurement outcome. We denote $x$, $y$ and $z$ the inputs of Alice, Bob and Charlie respectively, and $a$, $b$ and $c$ their outputs. The resulting statistics are given by:

\begin{equation} \label{quantum}
    p(abc|xyz) = \tr \left[ \left( M^A_{a|x} \otimes M^{B_1B_2}_{b|y} \otimes  M^C_{c|z} \right)  \left( \rho_{AB_1}  \otimes \rho_{B_2C} \right) \right]
\end{equation}
where $M^A_{a|x}$ denotes the operators of Alice's measurements (i.e. the POVM elements), satisfying $M^A_{a|x} \geq 0$ and $ \sum_a M^A_{a|x} = \openone$ for all $x$, and similarly for Bob's and Charlie's measurements. Note that Bob's measurement acts on both subsystems $B_1$ and $B_2$. Also when computing \eqref{quantum}, one should carefully associate the Hilbert space of the various subsystems. 

Given specific quantum correlations, one may now ask whether they could be reproduced by a classical model. The standard approach in nonlocality consists in asking whether the distribution admits a decomposition of the form: 
\begin{equation} \label{local}
     p(abc|xyz) = \int d \lambda p(\lambda)  p(a|x\lambda) p(b|y\lambda) p(c|z\lambda)
\end{equation}
where $\lambda$ denotes the shared classical variable distributed to all parties (Alice, Bob and Charlie) with distributions $p(\lambda)$. This represents the usual definition of Bell locality for a tripartite system, see e.g. \cite{review}.

At this point it is important to realize that comparing quantum distributions of the form \eqref{quantum} to classical correlations of the form \eqref{local} is arguably not appropriate. The key point is that the underlying network structure is not the same. Indeed, the decomposition in Eq. \eqref{local} assumes a common shared variable distributed to all parties; hence requiring a common source. However, this structure does not match the quantum network in Fig.~\ref{fig:bilocality}. 

This observation motivated the authors of Ref. \cite{Branciard_2010} to propose an alternative definition of locality for this problem, taking into account the network structure as in Fig.~\ref{fig:bilocality}. Indeed, one should now consider a local model involving two separate sources and enforce their independence. Just like the quantum state has a product structure, i.e. $\rho_{AB_1}  \otimes \rho_{B_2C}$, the classical model should feature the same structure. This motivates the notion of ``bilocal'' correlations, which take the form:
\begin{equation} \label{bilocal}
     p(abc|xyz) = \int d \lambda p(\lambda) \int d \mu q(\mu) p(a|x\lambda) p(b|y\lambda\mu) p(c|z\mu)
\end{equation}
where $\lambda$ and $\mu$ denote the two shared variables. Alice and Bob have access to $\lambda$ (distributed by the source $S_{AB}$), while Bob and Charlie have access to $\mu$ (via source $S_{BC}$). 
Importantly, $\lambda$ and
$\mu$ are distributed according to independent probability distributions $p(\lambda)$ and $q(\mu)$. 

There exist quantum correlations of the form \eqref{quantum} that are non bilocal, i.e. do not admit a decomposition of the form of Eq. \eqref{bilocal}. A simple (perhaps almost trivial) example consists in setting the state $\rho_{AB_1}$ to be a maximally entangled two-qubit Bell state, and choosing the measurement of Alice to be anti-commuting Pauli operators. Choosing Bob's measurement to be some appropriate anticommuting Pauli operators on the $B_1$ subsystem, and trivial on $B_2$, the resulting distribution $p(abc|xyz)$ (as given in \eqref{quantum}) will be non-bilocal, no matter what state $\rho_{B_2C}$ is produced by the second source and what measurements Charlie performs. This is because, the marginal distribution $p(ab|xy) = \sum_c p(abc|xyz)$ violates the standard CHSH Bell inequality. This deceptively simple example can be turned into a more elegant form following the construction of Fritz~\cite{Fritz_2012} by adding a third independent source between Alice and Charlie. This results in a scenario with two remarkable features: all inputs can be cleverly removed and replaced by outputs, hence resulting in the effect of ``quantum nonlocality without inputs'', and the obtained construction allows to easily demonstrate the presence of nonlocality in a network containing a loop. Nevertheless, one can still argue that this instance of network nonlocality has its origins in a standard Bell inequality test; see \cite{Fraser,Supic} and below.

Intuitively, however, quantum theory should allow for conceptually a very different (and possibly stronger) type of correlations in networks. This is because quantum theory allows not only for entangled states, but also for entangled measurements. Standard Bell tests exploit the entanglement of states, but usually do not involve entangled measurements. In contrast, the use of entangled measurements in networks is very natural. The combination of entangled states and entangled measurements offers the possibility to create strong correlations across the whole network. 

An interesting, and well-known, example is that of entanglement swapping. Take $\rho_{AB_1}$ and $\rho_{B_2C}$ to be two-qubit Bell states. Bob performs a joint measurement on both of his quantum subsystems, the so-called Bell-state measurement, projecting the two subsystems $B_1$ and $B_2$ in the basis of four maximally entangled Bell states. This has the remarkable effect of remotely preparing the systems of Alice and Charlie in an entangled state, even though these systems where initially fully independent from each other. By performing judicious local measurements, Alice and Charlie can then observe a Bell inequality violation (conditioned on each given output of Bob's measurement). This is known as ``event-ready nonlocality'' \cite{Zukowski}.

The question of characterizing this stronger form of quantum correlations proper to networks is our starting point in this work. Clearly one needs to go beyond the standard definition of bilocality (and its generalisation to arbitrary networks), as the above simple examples demonstrate. In a sense the question is to identify quantum correlations that truly take advantage of the network structure, and therefore do not reduce to standard Bell inequality tests. In the following, we take a first step in this direction and define a notion of genuine network nonlocality for quantum systems, which we illustrate with examples.

%



\section{Defining genuine network nonlocality}


\begin{figure*}[t]
    \centering
    \includegraphics[width = 0.9\textwidth, trim = 150 225 0 140]{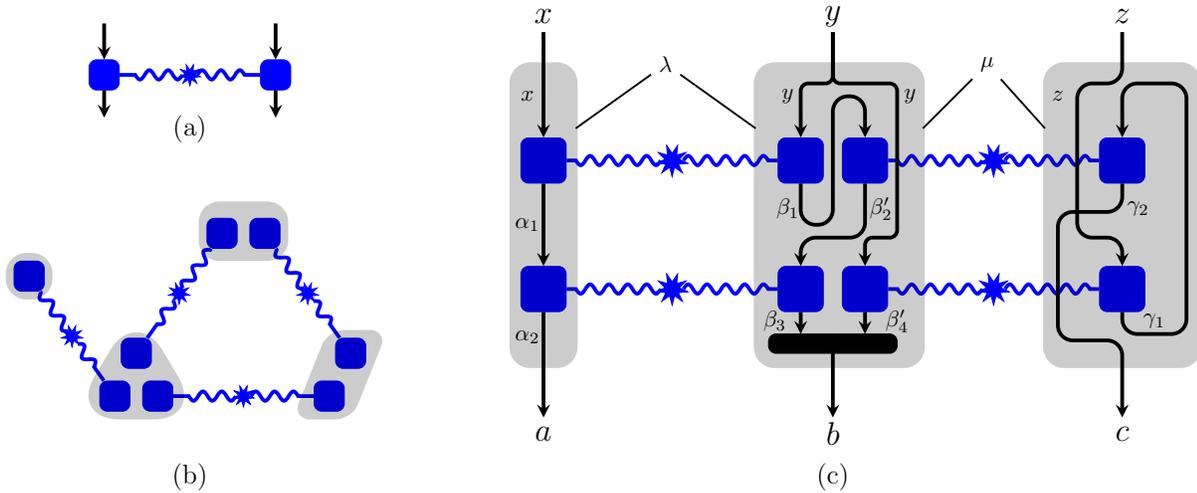}
    \caption{This figure explains the notion of quantum-wirable correlations. Quantum resources (states and measurements) are shown in blue, forming the quantum boxes, while classical information and processing is shown in black. (a) A bipartite quantum box is a device featuring two terminals, that receive inputs and produce outputs. The resulting correlations can be nonlocal, hence the quantum box represents the basic resource for standard quantum Bell nonlocality. (b) Quantum boxes distributed in a network: each party may have access to one or more terminals, performing classical processing on their inputs and outputs via a wiring. Correlations that can be achieved as such are termed ``quantum-wirable''. (c) An example of a quantum-wirable correlation in the bilocality network: given the shared variables ($\lambda$ and/or $\mu$) and their respective input, each party chooses a wiring of their terminals; this includes classical pre- and post-processing of the inputs and outputs, and the output of one terminal may be used as input to another. Notably, the order in which the terminals are used may be different for each party and may depend on the outputs of previous boxes.}
    \label{fig:qwirable}
\end{figure*}

To characterize genuine network nonlocality, we proceed by defining the opposite notion, namely correlations that are not genuine to a network, in the sense that they can be traced back to standard quantum Bell nonlocality. To do so, we ask ourselves how to use the resources of standard quantum Bell nonlocality in a network configuration. We take a rather resource-theoretic approach to this question, and view standard quantum nonlocality as ``quantum boxes'' that can be distributed by each source in the network to various subsets of parties (see Fig. \ref{fig:qwirable}). 

Let us first formalize the notion of a quantum box, which naturally follows from the standard scenario of Bell nonlocality. Consider here two distant parties (say Alice and Bob) sharing a quantum state $\sigma$. By performing local quantum measurements on their subsystems (represented by sets of POVMs $\{N_{\alpha|X}\}$ and $\{N_{\beta|Y}\}$, Alice and Bob can establish correlations of the form: 
\begin{equation}
    p(\alpha,\beta|X,Y) = \tr[(N_{\alpha|X} \otimes N_{\beta|Y} ) \sigma]
\end{equation}
where $X$ and $Y$ denote the inputs, and $\alpha$ and $\beta$ the outputs. We can view this probability distribution as a quantum box; it represents a (possibly nonlocal) box shared by Alice and Bob, which is realizable in quantum theory. Each party has access locally to one end of the box, i.e. a terminal, where they can give an input and then receive an output, see Fig.~\ref{fig:qwirable}\textcolor{myRed}{a}. Note these quantum boxes can be viewed as resources, and recent works have developed a resource-theoretic framework based on these ideas to characterize and quantify Bell nonlocality \cite{Wolfe2020quantifyingbell}; see also \cite{BarrettPR,Allcock} for previous works along these lines.

Coming back to the network configuration, we can now consider that  all sources in the network distribute quantum boxes to the parties, as shown in Fig.~\ref{fig:qwirable}\textcolor{myRed}{b}. Each party therefore receives a number of terminals which can be combined. As each terminal takes a classical input and delivers a classical output, the most general operation can be represented as a ``wiring'' of the terminals. This represents a classical measurement.

We now aim to characterize those correlations that can be established by using as resources quantum boxes distributed across the network, which are then locally combined via classical measurement, i.e. wirings. In the bilocality network (our focus here), this amounts to assuming that the first source distributes one (or more) quantum box(es) to Alice and Bob, and similarly for the second source connecting Bob and Charlie. Hence, each party now receives a number of terminals (each one featuring an input and an output), and can use all of them in order to produce an output given the received input, as depicted in Fig.~\ref{fig:qwirable}\textcolor{myRed}{c}. Upon receiving her input $x$, Alice chooses a wiring for processing her available terminals, in order to produce the final output $a$. For instance, in the figure, 
Alice receives two terminals and starts by inputting $x$ in the first one; then she uses the output of the first terminal as an input for the second one; the final output $a$ is then be chosen as the output of the second terminal.

Notably, these wiring operation can have non-trivial effects on nonlocality, in particular when applied to boxes that are nonlocal, allowing for instance to increase (or distill) nonlocality \cite{forster_2009,Brunner2009}. More generally, wiring operations feature a rich structure, in particular when three or more terminals are available; see e.g. \cite{ShortNJP,Gallego2017}. For instance, the output of the first terminal may then be used to decide on the order in which the next two terminals will be used. Even more possibilities arise, when considering a party that receives terminals from more than one source. This is the case for the middle party, Bob, in the bilocality network. At this point, it is also worth noting that it is in general not possible to contract all quantum boxes shared by two parties into a single box. This is because the order in which the boxes are used on one side might be different than the order used on the other side.

We are now ready to state our main definition. We introduce the class of ``quantum wirable'' correlations in a network as those that can be obtained when all the sources distribute quantum boxes to the parties, and the parties perform arbitrary local wiring operations. Formally, such distributions admit a decomposition of the form (here stated for the bilocality network): 
\begin{align}
\label{Qwirable}
    p(a,b,c|x,y,z) & =  \\
    \sum_{\substack{\lambda,\mu, \vec{k}, \\ \vec{l},\vec{l}',\vec{m},\\ \vec{X},\vec{Y},\vec{Z},\\ \vec{\alpha},\vec{\beta},\vec{\gamma}}}\ & \overline W_A(a,\vec{k},\vec{X}|x,\lambda)\, \overline W_B(b,\vec{l},\vec{l}',\vec{Y}|y,\lambda,\mu)  \nonumber \\[-4em]
    & \ \times \overline W_C(c,\vec{m},\vec{Z}|z,\mu)\, p(\lambda)\, p(\mu) \nonumber \\
    & \ \times p_Q^{\vec{k},\vec{l}}(\vec{\alpha},\vec{\beta}|\vec{X},\vec{Y},\lambda)\, p_Q^{\vec{l}',\vec{m}}(\vec{\beta}',\vec{\gamma}|\vec{Y},\vec{Z},\mu) \nonumber
\end{align}
where
\begin{equation}
    \overline W_A(a,\vec{k},\vec{X}|x,\lambda) = \prod_{i=1}^{n+1} W_A^i(k_i,X_i|[X]_{i-1},[\alpha]_{i-1})
\end{equation}
with the notation $[k]_i=\{k_0,k_1,\ldots,k_i\}$, $X_0=x$, $\alpha_0=\lambda$, $k_{n+1}=0$ and $a=X_{n+1}$ describes the successive wirings of Alice's $n$ boxes (similarly for $\overline W_B$ and $\overline W_C$), and $p_Q^{\vec{k},\vec{l}}(\vec{\alpha},\vec{\beta}|\vec{X},\vec{Y},\lambda)$ (and similarly $p_Q^{\vec{l}',\vec{m}}(\vec{\beta}',\vec{\gamma}|\vec{Y},\vec{Z},\mu)$) are distributions of multiple quantum boxes with interactions ordered by $\vec{k}$ and $\vec{l}$:
\begin{equation}
    p_Q^{\vec{k},\vec{l}}(\vec{\alpha},\vec{\beta}|\vec{X},\vec{Y},\lambda) = \prod_{i=1}^n p_Q(\alpha_{k^{-1}_i},\beta_{l^{-1}_i}|X_{k^{-1}_i},Y_{l^{-1}_i}).
\end{equation}
Here, $k$ and $k^{-1}$ (similarly for $l$, $l'$ and $m$) satisfy $k^{-1}_i\in\{0,\ldots,n\}$, $k_{k^{-1}_i}=i$, and $l_i l_i'=0$ for all $i=1,\ldots,n$.

As argued above, such distribution use as main resource the quantum boxes (which are possibly nonlocal). The local wiring operations are basically classical. One may thus argue that such distributions do not leverage the structure of the network to extend the nonlocality, and thus that they can be traced back to standard Bell nonlocality.

In our discussion above, we have already encountered an example of quantum non-bilocal correlations that are however quantum-wirable. The Fritz example is also clearly quantum-wirable; see Appendix A. 

We have also already seen examples of quantum correlations that are provably not quantum-wirable, i.e. do not admit a decomposition of the form \eqref{Qwirable}. This is the case for the entanglement swapping experiment discussed above. Indeed, the observation of event-ready nonlocality rules out any quantum-wirable model. This is because for any distribution of the form of \eqref{Qwirable}, the conditioned marginal distribution $p(ac|xyzb)$ is necessarily local, i.e. cannot violate any Bell inequality. More generally, the use of a wiring at Bob implies that, for every measurement $y$ and outcome $b$, the resulting quantum state held by Alice and Charlie must be separable, even though all the quantum boxes in the network may feature entangled states. Below we present a more interesting example of genuine network quantum nonlocality, based on the concept of self-testing. 


\section{Network self-testing}

Self-testing refers to the characterization of quantum systems in a black-box scenario \cite{MayersYao,SupicBowles}. In its standard form, self-testing involves a quantum Bell experiment. The observation of specific quantum nonlocal correlations (or maximal violation of some Bell inequality) can then be used to precisely certify the underlying entangled state and local measurements, up to local isometries. 

Here we present an instance of self-testing in the bilocality network. The interesting feature of this example is that the involved correlations are non-bilocal, but are local following the usual definition of tripartite Bell locality (see below). Hence, this represents an example of genuine network quantum nonlocality which does not rely on event-ready nonlocality.

Consider both sources producing a Bell pair, i.e. $\rho_{AB_1} = \ketbra{\phi_+}{\phi_+}$ and $\rho_{B_2C} = \ketbra{\phi_+}{\phi_+}$. Alice and Charlie perform the same local measurements, represented by the Pauli observable $A_0 = C_0 = \frac{\sigma_\mathsf{x}+\sigma_\mathsf{z}}{\sqrt{2}}$ and $A_1 = C_1 = \frac{\sigma_\mathsf{x}-\sigma_\mathsf{z}}{\sqrt{2}}$. Bob applies the Bell state measurement. With these resources the parties observe correlations $p(abc|xz)$ with the following properties:
\begin{align} \label{pB}
     \sum_{ac} p(abc|xz)  
    =\tr\left[\left(\idd^{A}\tp M_{b}^{B_1B_2}\tp\idd^{C}\right)\left(\rho_{AB_1}\tp \rho_{B_2C}\right)\right] = \frac{1}{4}, 
\end{align}
for all $b$, where $b$ can have four different values, and thus can be represented as two bits $b = (b_1,b_2)$. Let us define the state of Alice and Charlie remotely prepared by Bob's measurement:
\begin{equation}
\label{rho_b}
\varrho_{b} = 
\frac{\tr_{B_1B_2}\left[\left(\idd^{A}\tp M_{b}^{B_1B_2}\tp\idd^{C}\right)\left(\rho_{AB_1}\tp \rho_{B_2C}\right)\right]}{\tr\left[\left(\idd^{A}\tp M_{b}^{B_1B_2}\tp\idd^{C}\right)\left(\rho_{AB_1}\tp \rho_{B_2C}\right)\right]}
\end{equation}
%
%
The statistics of Alice and Charlie, conditioned on Bob's output, are given by:
\begin{align}\label{marg1}
\tr\left[(A_0 \tp \idd)\varrho_{b}\right] &= \tr\left[(A_1 \tp \idd)\varrho_{b}\right] = 0,\\ \label{marg2} \tr\left[(\idd \tp C_0)\varrho_{b}\right] &= \tr\left[(\idd \tp C_1)\varrho_{b}\right] = 0,\\
\label{rhogen1}
\tr\left[(A_0 \tp C_0)\varrho_{b}\right] &= \tr\left[(A_1 \tp C_1)\varrho_{b}\right] = (-1)^{b_1}\delta_{b_1b_2},  \\ \label{rhogen2}
\tr\left[(A_0 \tp C_1)\varrho_{b}\right] &= \tr\left[(A_1 \tp C_0)\varrho_{b}\right] = (-1)^{b_1}\delta_{b_11-b_2},
\end{align}
where $\delta_{jk}$ is the Kronecker delta. 
We are now in position to state our self-testing result:

\begin{thm}\label{lemma1}
The correlations in Eqs \eqref{pB} and \eqref{marg1}-\eqref{rhogen2} imply that (i) the two local measurements of both Alice and Charlie are anticommuting, (ii) the measurement performed by Bob is the Bell-state measurement, and (iii), all states  $\varrho_b$ (as in Eq. \eqref{rho_b}) are Bell states, i.e. Bob's measurement remotely steers maximally entangled state for Alice and Charlie.
\end{thm}

We now sketch the proof; the full proof is given in Appendix \ref{lemmaproof}. First, from Jordan's lemma, we can bring Alice's and Charlie's measurement observables to a block-diagonal form. A block diagonal unitary transformation can rotate Alice's observable to a form where $A_0$ has in every block Pauli's $\sz$, while all blocks of $A_1$  are rotated to the real plane of the Bloch sphere. There exists a unitary making the equivalent rotation of Charlie's measurement observables. This allows to reduce the problem to an ensemble of qubit states. The correlations given in eqs. \eqref{marg1}-\eqref{rhogen2} imply that one of the two following statements must hold: either the pairs of measurements of Alice and Bob commute, or they anticommute. The first possibility is ruled out by using the fact that the measurement correlations $\{p(abc|xz)\}$ are non-bilocal. Hence, the observables of Alice and Charlie are unitarily equivalent to $\sz\otimes\idd$ and $\sx\otimes\idd$. With this insight \eqref{marg1}-\eqref{rhogen2} imply that for every output $b$ the state $\varrho_b$ corresponds to a Bell pair in tensor product with some junk state (on which the measurements of Alice and Charlie act trivially). By tracing out the uncorrelated degrees of freedom we get a quantum channel mapping the steered states $\varrho_b$ to qubit Bell pairs. Finally, this insight together with the argument presented in \cite{renou2018self} concludes the self-testing of Bob's measurement.

As mentioned above, an interesting feature of this self-testing result is that it relies on the fact that the distribution is not bilocal. Indeed, one can readily check that the quantum distribution is Bell local in the usual sense, i.e. it admits a decomposition of the form \eqref{local}. This in stark contrast to previous self-testing results in networks \cite{renou2018self,JD_BSM} (and possibly with all known self-testing results) which rely on standard Bell inequality violation. 

 Moreover, as the measurement performed by Bob must be the Bell-state measurement, remotely preparing maximally entangled states $\varrho_b$ for Alice and Charlie, it is clear that the quantum distribution is genuine network nonlocal. But since the distribution is Bell local it does not feature event-ready nonlocality.

%

\section{Discussion and outlook}

We have defined a notion of genuine network quantum  nonlocality. This represents a stronger form of quantum nonlocality in networks (compared to the previous notion of non-bilocality), which can only be obtained from the judicious combination of entangled states and non-classical measurements. Therefore, the nonlocality of such correlations cannot be traced back to standard Bell inequality violations, and represents a phenomenon that is proper to the network configuration. We presented an instance of this effect based on self-testing in the bilocality network. Notably, this can be seen as a novel form of self-testing, proper to networks, as it relies entirely on non-bilocality, and does not involve the violation of a standard Bell inequality.

Our approach (in particular the definition in Eq. \eqref{Qwirable}) can be readily generalized to more complex networks. A case of particular interest is the triangle network. We conjecture that the distribution recently proposed by Renou et al. \cite{renou_2019} is genuine network nonlocal in the sense that we define, contrary to the distribution of Fritz \cite{Fritz_2012} which is quantum-wirable.

An interesting question is to identify the necessary quantum resources for demonstrating genuine network nonlocality. From our examples, one may think that the presence of an entangled measurement is necessary. However, this may not be the case in general. Indeed, from the definition of quantum wirable correlations in Eq. \eqref{Qwirable}, we see that genuine network nonlocality requires the use of a non-classical measurement, i.e. not describable as a wiring. Interestingly there exist quantum  measurement that feature only separable eigenstates, but yet are non-classical \cite{Bennett}; see also \cite{VertesiMiguel}. This effect is known as ``quanutm nonlocality without entanglement'', It would be interesting to see if the use of such a measurement could lead to genuine network nonlocality.

It is worth pointing out that our approach is completely different from the one taken in a series of recent works \cite{MiguelElie,Chao,Bierhorst,Coiteux}. The latter address a different question, namely they aim at characterizing multipartite quantum states (such as GHZ states) and their correlations which cannot be obtained from unbounded bipartite resources (quantum or even post-quantum)  distributed in a triangle network. In contrast, our work focuses on quantum correlations that are possible in a given network structure, but which cannot be decomposed using quantum boxes and local wirings.

Finally, note that one could consider a natural variation of our definition of quantum-wirable correlations to ``no-signaling-wirable'' correlations. That is, one would demand a decomposition as in Eq. \eqref{Qwirable}, but where the shared nonlocal boxes would no longer need to admit a quantum realization, but only satisfy the no-signaling constraints (so-called non-signalling or Popescu-Rohrlich boxes \cite{PR,BarrettPR}). Note that any instance of event-ready quantum nonlocality is then already an example of a quantum distribution that is not no-signaling-wirable \footnote{Similarly to the quantum case discussed in the main text, the use of a wiring by Bob implies that the remotely prepared correlations for Alice and Charlie must be local; see e.g. \cite{ShortNJP}}. It would be interesting to see if this is also the case for our self-testing example.

\emph{Note added.} In a recent pre-print Pozas-Kerstjens et al. \cite{pozas} propose a notion of ``full network nonlocality'' which is different and complementary to ours. Full network nonlocality, which certifies that all sources in the network produce an entangled state, does not imply genuine network nonlocality (as we define); indeed the presence of entanglement on all links can be enforced via standard Bell tests (e.g. CHSH), which results in correlations that are quantum-wirable.

\begin{acknowledgements}
  This work was supported by the Swiss National Science Foundation (project 2000021 192244/1 and NCCR SwissMap).
\end{acknowledgements}

\bibliographystyle{unsrturl}
\bibliography{biblio}

\newpage

\appendix

\section{The Fritz distribution is quantum-wirable}
\begin{figure}[t]
    \centering
    \includegraphics[scale=0.7]{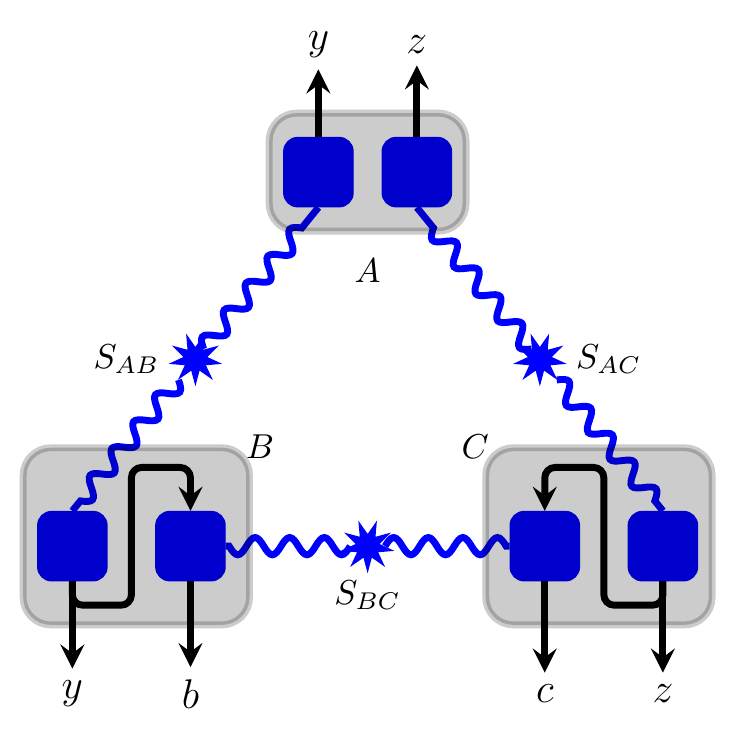}
    \caption{The underlying quantum network of Fritz's distribution. The distribution is network nonlocal but quantum-wirable.}
    \label{fig:fritz}
\end{figure}

In~\cite{Fritz_2012}, Theorem 2.16, Fritz introduced a nonlocal distribution in the triangle network based on CHSH violation. In the triangle network, as shown in Fig.~\ref{fig:fritz}, two sources $S_{AB}$ and $S_{AC}$ send fully correlated classical bits, while $S_{BC}$ sends the maximally entangled states. Alice outputs the two bits that she receives, denoted as $(y,z)\equiv A$. Bob (Charlie) performs the optimal CHSH measurements, treating $y$ from $S_{AB}$ ($z$ from $S_{AC}$) as the input setting, and outputs two bits $(y,b)\equiv B$ ($(z,c)\equiv C$ respectively). The Bell nonlocality of $p(bc|yz)$ implies the network nonlocality of $p(ABC)$.

However, we can clearly see that this distribution is quantum-wirable, since all measurements are separable and the output are classical processing of output of quantum boxes.

\section{Proof of Theorem \ref{lemma1}}\label{lemmaproof}

Jordan's lemma states that for any two dichotomic observables there exist a unitary which transforms both of them to a block diagonal form in the same basis, where the blocks are of size $2\times 2$ or $1\times 1$. In the same way like in \cite{Jed}, we embed every $1\times 1$ into a Hilbert space of larger dimension. This operation does not affect the correlation probabilities, and it simplifies our analysis, as we work with a Jordan decomposition in which all blocks are of the size $2\times 2$. Henceforth, through a simple basis change, we can bring Alice's and Charlie's measurement observables to a block-diagonal form
\begin{align}
    V_AA_0V_A^\dagger &= \bigoplus_{\alpha}\sigma_\mathsf{z}^{\alpha},\\ V_AA_1V_A^\dagger &= \bigoplus_\alpha \left(\cos\theta_\alpha\sigma_\mathsf{z}^\alpha + \sin\theta_\alpha\sigma_\mathsf{x}^\alpha\right),\\
    V_CC_0V_C^\dagger &= \bigoplus_{\gamma}\sigma_\mathsf{z}^{\gamma},\\ V_CC_1V_C^\dagger &= \bigoplus_\gamma \left(\cos\phi_\gamma\sigma_\mathsf{z}^\gamma + \sin\phi_\gamma\sigma_\mathsf{x}^\gamma\right),
\end{align}
where $\theta_\al,\phi_\gm \in [0,2\pi)$ for all $\al$ and $\gm$. In turn, this implies that we can assume that
\begin{equation}
    V\varrho_{b_1b_2}V^\dagger = \bigoplus_{\alpha,\gamma}p_{b_1b_2}^{\alpha,\gamma}\varrho_{b_1b_2}^{\alpha,\gamma},
\end{equation}
where $V = V_A\otimes V_C$, without loss of generality.

The correlations between $A_0$ and $C_0$ imply
\begin{align}
    \tr\left[\left(A_0\tp C_0\right)\varrho_{00}\right] &= \sum_{\alpha,\gamma}p_{00}^{\alpha,\gamma}\tr\left[\left(\sigma_\mathsf{z}^\alpha\tp \sigma_\mathsf{z}^\gamma\right)\varrho_{00}^{\alpha,\gamma}\right]\\
    &= 1
\end{align}
Given that $\sum_{\alpha,\gamma}p_{00}^{\alpha,\gamma} = 1$ the last equation implies $\tr\left[\left(\sigma_\mathsf{z}^\alpha\tp \sigma_\mathsf{z}^\gamma\right)\varrho_0^{\alpha,\gamma}\right] = 1$ for all $\alpha$ and $\gamma$ such that $p_{00} \neq 0$ (components of $\varrho_{00}$ with $p_{00} = 0$ are not relevant because they do not contribute to the observed statistics). This condition imposes the following structure of blocks on the diagonal of $V\varrho_{00}V^\dagger$:
\begin{equation}
    \varrho_{00}^{\alpha,\gamma} = \begin{bmatrix}
    q_{00}^{\al,\gm} & 0 & 0 & r_{00}^{\al,\gm}\\
    0 & 0 & 0 & 0\\
    0 & 0 & 0 & 0\\
    {r_{00}^{\al,\gm}}^* & 0 & 0 & 1-q_{00}^{\al,\gm}
    \end{bmatrix}.
\end{equation}
Given this form of $\varrho_{00}$ and the fact that $\tr[\sigma_\mathsf{x}^\alpha\tp\sigma_\mathsf{z}^\gamma\rho_{00}^{\alpha,\gamma}] = \tr[\sigma_\mathsf{z}^\alpha\tp\sigma_\mathsf{x}^\gamma\rho_{00}^{\alpha,\gamma}] = 0$ for all $\al$ and $\gm$ with $p_{00}\neq 0$, the correlations between $A_1$ and $C_1$ imply
\begin{align} \nonumber
    &\tr\left[\left(A_1\tp C_1\right)\varrho_{00}\right] = \\ \nonumber &=\sum_{\alpha,\gamma}p_{00}^{\alpha,\gamma}\tr\big[\big(\left(\cos\theta_\alpha\sigma_\mathsf{z}^\alpha + \sin\theta_\alpha\sigma_\mathsf{x}^\alpha\right)\tp \\ &\qquad\qquad\tp\left(\cos\phi_\gamma\sigma_\mathsf{z}^\gamma + \sin\phi_\gamma\sigma_\mathsf{x}^\gamma\right)\big)\varrho_{00}^{\alpha,\gamma}\big]\\ \nonumber
    &= \sum_{\alpha,\gamma}p_{00}^{\alpha,\gamma}\left[\cos\theta_\alpha\cos\phi_\gamma + \sin\theta_\alpha\sin\phi_\gamma 2\text{Re}(r_{00}^{\al,\gm})\right]\\ \label{11}
    &= 1.
\end{align}
The positivity of the blocks imposes $|r_{00}^{\al,\gm}|^2 \leq q_{00}^{\al,\gm}(1-q_{00}^{\al,\gm})$, which further implies $|r_{00}^{\al,\gm}| \leq 1/2$ and therefore $2|\text{Re}(r_{00}^{\al,\gm})| \leq 1$. Hence, the maximal value of the expression $\cos\theta_\alpha\cos\phi_\gamma + \sin\theta_\alpha\sin\phi_\gamma 2\text{Re}(r_{00}^{\al,\gm})$ for all $\alpha$ and $\gamma$ such that $p_{00}\neq 0$ is equal to $1$. The last equation thus implies that for every $\alpha$ and $\gamma$ with $p_{00}\neq 0$ this expression must take value $1$ since $\sum_{\alpha,\gamma}p_{00}^{\alpha,\gamma} = 1$. For every $\al$ and $\gm$ such that $p_{00}^{\al,\gm} \neq 0$, the equation
\begin{equation}
    \cos\theta_\alpha\cos\phi_\gamma + \sin\theta_\alpha\sin\phi_\gamma 2\text{Re}(r_{00}^{\al,\gm}) = 1
\end{equation}
admits two groups of solutions :
\begin{enumerate}
    \item $2\text{Re}(r_{00}^{\al,\gm}) =  1$ and $\theta_\alpha = \phi_\gamma \neq \pi$ or $2\text{Re}(r_{00}^{\al,\gm}) =  -1$ and $\theta_\alpha = 2\pi - \phi_\gamma \neq \pi$;
    \item $\theta_\alpha = \phi_\gamma = \pi$ and unrestricted value of $r_{00}^{\al,\gm}$.
\end{enumerate}
Observing the correlations between $A_i$ and $C_j$ for Bob's outcome $b = {11}$ we come to similar conclusions, i.e, two groups of solutions exist which satisfy all reference correlations for all $\alpha$ and $\gamma$ such that $p_{11}^{\alpha,\gamma}\neq 0$:
\begin{enumerate}
    \item $2\text{Re}(r_{11}^{\al,\gm}) =  -1$ and $\theta_\alpha = \phi_\gamma \neq l\pi$ or $2\text{Re}(r_{11}^{\al,\gm}) =  1$ and $\theta_\alpha = 2\pi - \phi_\gamma \neq \pi$ ;
    \item $\theta_\alpha = \phi_\gamma = \pi$ and unrestricted value of $r_{11}^{\al,\gm}$.
\end{enumerate}
Observe now that the unitary $U = \oplus_\gamma U_\gamma$, where $U_\gamma$ is
\begin{equation}
    U_\gamma = \begin{bmatrix}
    \cos(\frac{\phi_\gamma}{2}) & \sin(\frac{\phi_\gamma}{2})\\
    -\sin(\frac{\phi_\gamma}{2}) & \cos(\frac{\phi_\gamma}{2})
    \end{bmatrix},
\end{equation}
induces the following map $\tilde{C}_0 = UV_CC_1V_C^\dagger U^\dagger = \bigoplus_\gamma\sigma_\mathsf{z}^\gamma$ and $\tilde{C}_1 = UV_CC_0V_C^\dagger U^\dagger = \bigoplus_\gamma(\cos(\phi_\gamma)\sigma_\mathsf{z}^\gamma - \sin(\phi_\gamma)\sigma_{\mathsf{x}}^\gamma)$. With this local change of Charlie's basis the correlations for Bob's outcome $b={01}$ are
\begin{align}\nonumber
\tr[A_0\otimes\tilde{C}_0\tilde{\varrho}_{01}] &= 1\\ \nonumber
\tr[A_1\otimes\tilde{C}_1\tilde{\varrho}_{01}] &= 1\\  \nonumber
\tr[A_0\otimes\tilde{C}_1\tilde{\varrho}_{01}] &= 0\\ \label{2704}
\tr[A_1\otimes\tilde{C}_0\tilde{\varrho}_{01}] &= 0
\end{align}
where $\tilde{\varrho}_{01} = (\idd\otimes U)V\varrho_{01}V^\dagger(\idd\otimes U^\dagger)$. The first equation \eqref{2704} implies %
\begin{equation}
    \tilde{\varrho}_{01}^{\alpha,\gamma} = \begin{bmatrix}
    {\tilde{q}}_{01}^{\al,\gm} & 0 & 0 & {\tilde{r}}_{01}^{\al,\gm}\\
    0 & 0 & 0 & 0\\
    0 & 0 & 0 & 0\\
    {\tilde{r}}_{01}^{{\al,\gm}^*} & 0 & 0 & 1-{\tilde{q}}_{01}^{\al,\gm}
    \end{bmatrix},
\end{equation}
where $\tilde{\varrho}_{01}^{\alpha,\gamma} = (\idd\otimes U)V\varrho_{01}^{\alpha,\gamma}V^\dagger(\idd\otimes U^\dagger)$.
Writing explicitly all the correlations, analogously to the previous cases, we again obtain two groups of solutions for all $\alpha$ and $\gamma$ such that $p_{01}^{\alpha,\gamma}\neq 0$:
\begin{enumerate}
    \item $2\text{Re}({\tilde{r}}_{01}^{\al,\gm}) =  1$ and $\theta_\alpha = 2\pi - \phi_\gamma \neq \pi$ or $2\text{Re}({\tilde{r}}_{01}^{\al,\gm}) =  -1$ and $\theta_\alpha =  \phi_\gamma \neq \pi$;
    \item $\theta_\alpha = \phi_\gamma = \pi$  and unrestricted value of ${\tilde{r}}_{01}^{\al,\gm}$.
\end{enumerate}
Equivalently, for Bob's output $b={10}$ we define $\tilde{\varrho}_{10} = (\idd\otimes U)V\varrho_{10}V^\dagger(\idd\otimes U^\dagger)$ and obtain two groups of solutions for all $\alpha$ and $\gamma$ such that $p_{10}^{\alpha,\gamma}\neq 0$:
\begin{enumerate}
    \item $2\text{Re}({\tilde{r}}_{10}^{\al,\gm}) =  1$ and $\theta_\alpha =  \phi_\gamma \neq \pi$ or $2\text{Re}({\tilde{r}}_{10}^{\al,\gm}) =  -1$ and $\theta_\alpha = 2\pi - \phi_\gamma \neq \pi$;
    \item $\theta_\alpha = \phi_\gamma = \pi$ and unrestricted value of ${\tilde{r}}_{10}^{\al,\gm}$.
\end{enumerate}

Let us assume the second case solution holds for all outputs $b$, \emph{i.e.} for all $\al$, $\gm$ and $b$ such that $p_{b}^{\alpha,\gamma}\neq 0$. This implies $\cos\theta_\alpha = \cos\phi_\gamma = \pm 1$ for all $\alpha$ and $\gamma$ such that $p_b^{\al,\gm}\neq 0$ for either of $b = 00,11,01,10$. In this case measurement observables $A_0$ and $A_1$ are commuting on the support of all states $\varrho_b$, as well as $C_0$ and $C_1$. 

Suppose Alice and Charlie both perform commuting measurements. The probability distribution $p(a,b,c|x,z)$ is then such that one can write a parent distribution $p(a_0,a_1,b,c_0,c_1)$ without inputs, where $a_x$ denotes the output $a$ for the setting $x$, and similarly for $c_z$ \cite{fine1982hidden}. The distribution $p(a,b,c|x,z)$ can then be obtained as a marginal of the parent distribution, e.g. by selecting the desired outcome $a_x$ as a function of the input $x$.

It has been proven by Branciard \emph{et al.} \cite{branciard2012bilocal}, that one cannot generate a non-bilocal distribution in this scenario if Bob has no inputs. We present the argument here. Take an arbitrary distribution $p(a,b,c)$. Due to the bilocality assumption, it can be decomposed as $p(a,b,c)=p(a)p(c)p(b|a,c)$, where we used that $p(c|a)=p(c)$. Now consider that the source between Alice and Bob samples $\lambda$ from Alice's marginal distribution $p(a)$ and Alice outputs $a= \lambda$. Similarly, the source between Bob and Charlie samples $\mu$ from Charlie's marginal distribution $p(c)$  and Charlie outputs $c= \mu$. Then upon receiving $\lambda$ and $\mu$, Bob can produce an output according to $p(b|a=\lambda, c=\mu)$, hence reproducing the arbitrary distribution $p(a,b,c)$. Therefore, any local distribution admits a bilocal model in absence of inputs for Bob.

Thus we conclude that if Alice and Charlie's measurements commute, then the resulting statistics admits a bilocal model, and can therefore not violate any bilocality inequality. Since the correlations we consider maximally violate the bilocality inequality from \cite{Branciard_2010} the possibility that both Alice and Charlie perform compatible measurements must be discarded.

Therefore, there must be $\alpha$ and $\gamma$ such that $r_b^{\alpha,\gamma} = \pm 1$ (or $\tilde{r}_b^{\alpha,\gamma} = \pm 1$). In that case the block $\varrho_b^{\alpha,\gamma}$ is maximally entangled, and thus the full state $\varrho_b$ is entangled. 


We denote the set of values of $\al$ such that there is $\gamma$ such that $\cos(\theta_\al)\cos(\phi_\gm) = 1$ with $\mathcal{A}'$. Similarly define the set of values of $\gm$ such that there is $\al$ such that $\cos(\theta_\al)\cos(\phi_\gm) = 1$ with $\mathcal{G}'$. The sets of all other values of $\al$ and $\gm$ is denoted with $\mathcal{A}''$ and $\mathcal{G}''$, respectively. Without loss of generality assume there is $p_{00}^{\al,\gm} \neq 0$ for $\al \in \mathcal{A}''$ and $\gm \in \mathcal{G}''$ such that $\cos(\theta_\al)\cos(\phi_\gm) \neq 1$. There are then two possibilities, namely $\theta_\alpha = \phi_\gamma \neq \pi$ implying $2r_{00}^{\al,\gm} = 1$ and $\theta_\alpha = 2\pi - \phi_\gamma \neq \pi$ implying $2r_{00}^{\al,\gm} = -1$. For the time being, we chose the first one, and remember that it might as well have been the other one, implying that $\varrho_{00}^{\al\gm} = \Phi_-$. We deal with this other possibility later.

We thus have $2\text{Re}(r_{00}^{\al,\gm}) = 1$, which implies  also $q_{00}^{\al,\gm} = 1/2$. Hence, 
\begin{equation}
\varrho_{00}^{\al\gm} = \Phi_+,
\end{equation}
for $\al \in \mathcal{A}''$ and $\gm \in \mathcal{G}''$.
At this stage, we also notice that eq. \eqref{11} can only hold if $\al \in \mathcal{A}'$ and $\gm \in \mathcal{G}''$ or $\al \in \mathcal{A}''$ and $\gm \in \mathcal{G}'$ implies $p_{00}^{\al,\gm} = 0$. Furthermore when $\al \in \mathcal{A}'$ and $\gm \in \mathcal{G}'$, $\theta_\al \neq \phi_\gm$ only if $p_{00}^{\al,\gm} = 0$.

Similarly we see that the correlations between $A_0$ and $C_0$, and $A_1$ and $C_1$ allow to conclude that the state $\varrho_{11}$ has the following form $\varrho_{11} = \oplus p_{11}^{\al\gm}\varrho_{11}^{\al\gm}$ where for $\al \in \mathcal{A}''$ and $\gm \in \mathcal{G}''$
\begin{equation}
    \varrho_{11}^{\al\gm} =\Psi_{-}.
\end{equation}
For all the cases when $\theta_\al = 2\pi - \phi_\gm$ we have $\varrho_{11}^{\al\gm} =\Psi_{+}$.
The analysis of the correlations of the states $\varrho_{01}$ and $\varrho_{10}$ can be done in an equivalent way. For $\al \in \mathcal{A}''$ and $\gm \in \mathcal{G}''$  we have
\begin{equation}
\tilde{\varrho}_{01}^{\alpha,\gamma} = \Phi_-.
\end{equation}
and if $\theta_\al = 2\pi-\phi_\gm$ then $\tilde{\varrho}_{01}^{\alpha,\gamma} = \Phi_+$.
For $\al \in \mathcal{A}'$ and $\gm \in \mathcal{G}''$ or $\al \in \mathcal{A}'$ and $\gm \in \mathcal{G}''$ it holds $p_{01}^{\al,\gm} = 0$.
By repeating the same procedure for the correlations on $\varrho_{10}$ we obtain that for $\al \in \mathcal{A}''$ and $\gm \in \mathcal{G}''$ we have
\begin{equation}
\tilde{\varrho}_{10}^{\alpha,\gamma} = \Psi_+.
\end{equation}
or if $\theta_\al = 2\pi-\phi_\gm$ then $\tilde{\varrho}_{10}^{\alpha,\gamma} = \Psi_-$, while for $\al \in \mathcal{A}'$ and $\gm \in \mathcal{G}''$ or $\al \in \mathcal{A}'$ and $\gm \in \mathcal{G}''$ it must be $p_{10}^{\al,\gm} = 0$. 
Let us now observe that 
\begin{equation}
    \tr_{\rB_1\rB_2}[\varrho^{\rA\rB_1}\otimes\varrho^{\rB_2\rC}] = \sum_bp(b)\varrho_b.
\end{equation}
On the left hands side we have the tensor product state $\varrho^\rA\otimes\varrho^\rC$, where $\varrho_\rA = \tr_{\rB_1}\left[\varrho^{\rA\rB_1}\right]$ and $\varrho_\rC = \tr_{\rB_2}\left[\varrho^{\rB_2\rC}\right]$. On the right hand side we have the state of the form 
\begin{equation}
    \bigoplus_{\al \in \mathcal{A'}, \gm\in\mathcal{G}'}p_{\al,\gm}'\varrho_{\al,\gm}' \bigoplus_{\al \in \mathcal{A''},\gm '\in \mathcal{G}''}p_{\al,\gm}''\varrho_{\al,\gm}''
\end{equation}
which has a tensor product structure only if $p'_{\alpha,\gamma} = 0$ for all $\al \in \mathcal{A}'$ and $\gm \in \mathcal{G}'$ or $p''_{\alpha,\gamma} = 0$ for all $\al \in \mathcal{A}''$ and $\gm \in \mathcal{G}''$. Given that the second case implies that Alice's and Charlie's measurements are compatible on the support of $\varrho_\rA$ and $\varrho_\rC$, respectively, for the reasons already explained this cannot be the case. A similar argument impedes having $\theta_\alpha \neq \phi_\gm$ when $\al \in \mathcal{A}''$ and $\gm \in \mathcal{G}''$, implying that for all $\al \in \mathcal{A}''$ and $\gm \in \mathcal{G}''$ it must hold $\theta_\al = \phi_\gm \equiv \theta$. This excludes the possibility $\theta_\al = 2\pi - \phi_\gm$, so we have to abandone this possibility. Hence, we conclude that Alice's and Charlie's measurements on the support of $\varrho_\rA$ and $\varrho_\rC$ respectively have the following form:
\begin{align}
    V_AA_1V_A^\dagger &= \bigoplus_\alpha(\cos(\theta)\sigma_\mathsf{z}^\alpha + \sin(\theta)\sigma_\mathsf{x}^\alpha)\\
    V_CC_1V_C^\dagger &= \bigoplus_\gamma(\cos(\theta)\sigma_\mathsf{z}^\gamma + \sin(\theta)\sigma_\mathsf{x}^\gamma)
\end{align}
Now we can return to the measurement statistics $\tr\left[A_0\otimes C_1\varrho_0\right] = 0$. This implies
\begin{equation}
    0 = \cos\theta\sum_{\al,\gm}p''_{\al,\gm} = \cos\theta,
\end{equation}
implying
\begin{align}\label{xmeasa}
    V_AA_1V_A^\dagger &= \bigoplus_\alpha\sigma_\mathsf{x}^\alpha\\
    &= \sigma_\mathsf{x} \otimes \idd \nonumber \\ \label{xmeasc}
    V_CC_1V_C^\dagger &= \bigoplus_\gamma\sigma_\mathsf{x}^\gamma\\
    &= \sigma_\mathsf{x} \otimes \idd \nonumber 
\end{align}
This implies that on the support of $\varrho_\rA$ and $\varrho_\rC$ Alice's and Charlie's measurements, respectively, anticommute. The corresponding Hilbert spaces $\mathcal{H}^A$ and $\mathcal{H}^C$ decompose: $\mathcal{H}^A = \mathcal{H}^{A'}\otimes \mathcal{H}^{A''}$ and $\mathcal{H}^C = \mathcal{H}^{C'}\otimes \mathcal{H}^{C''}$, where $\mathcal{H}^{A'}$ and $\mathcal{H}^{C'}$ are qubit Hilbert spaces. Finally we notice that since $\phi_\gm = \pi$ for all $\gm$ the unitary $U$ simply decomposes as $U = \tilde{U}\otimes \idd$, where 
\begin{equation}
    \tilde{U} = \frac{1}{\sqrt{2}}\begin{bmatrix}1 & 1\\-1&1\end{bmatrix}.
\end{equation}

Furthermore the states have the following form:
\begin{align}
    V\varrho_{00}V^\dagger &= \bigoplus_{\al,\gm}\Phi_+^{\al,\gm} = \Phi_+\otimes\sigma_{00},\\
    V\varrho_{11}V^\dagger &= \bigoplus_{\al,\gm}\Psi_-^{\al,\gm} = \Psi_-\otimes\sigma_{11},\\
    V{\varrho}_{01}V^\dagger &= (\idd\otimes U^\dagger)\bigoplus_{\al,\gm}\Phi_-^{\al,\gm}(\idd\otimes U)\\ &= (\idd\otimes U^\dagger)\Phi_-\otimes\sigma_{01} (\idd\otimes U),\\
    V{\varrho}_{10}V^\dagger &= (\idd\otimes U^\dagger)\bigoplus_{\al,\gm}\Psi_+^{\al,\gm}(\idd\otimes U)\\ &= (\idd\otimes U^\dagger)\Psi_+\otimes\sigma_{10}(\idd\otimes U),
\end{align}
where $\sigma_{jk}$ are junk states of additional degrees of freedom. Let us now define the CPTP map $\Lambda_A:\mathcal{H}^A\longrightarrow\mathcal{H}^{A'}$ such that $\Lambda_A(\rho) = \tr_{A''}\left[V_A\rho V_A^\dagger\right]$. Analogously we can define  $\Lambda_A:\mathcal{H}^C\longrightarrow\mathcal{H}^{C'}$ such that $\Lambda_C(\rho) = \tr_{C''}\left[V_C\rho V_C^\dagger\right]$. With these CPTP maps we have:
\begin{align}
    \Lambda_A\otimes\Lambda_C(\varrho_{00}) &=  \Phi_+,\\
    \Lambda_A\otimes\Lambda_C(\varrho_{11}) &=  \Psi_-,\\
    \Lambda_A\otimes\Lambda_C(\varrho_{01}) &=  (\idd\otimes \tilde{U}^\dagger)\Phi_- (\idd\otimes \tilde{U}),\\
    \Lambda_A\otimes\Lambda_C(\varrho_{10}) &=  (\idd\otimes \tilde{U}^\dagger)\Psi_+(\idd\otimes \tilde{U}),
\end{align}
These equations can be phrased as Step 1 in the proof of self-testing theorem of the Bell state measurement in \cite{renou2018self}. Since Step 2 which completes the proof depends completely on Step 1, we can use exactly the same methods to show that Bob's measurement can be self-tested to be exactly the Bell state measurement.
\end{document}